\documentclass[twoside]{article}
\usepackage{amsmath,amsfonts,amsthm}
\usepackage[affil-it]{authblk}

\usepackage{amstext}
\usepackage{amsgen}
\usepackage{amsbsy}
\usepackage{amsopn}
\usepackage{amssymb}
\usepackage{graphicx}
\usepackage{epsfig}

\newtheorem{theorema}{Theorem}
\newtheorem{algorithm}[theorema]{Algorithm}

\newtheorem{prop}{Theorem}
\newtheorem{proposition}[prop]{Proposition}

\newcommand{\Prob}{$\mathcal{P}$}
\newcommand{\SAT}{{SAT}$(n,m)$}
\newcommand{\SSAT}{{SSAT}$(n,m)$}

\def\IMAGESPATH{.}

\begin{document}

\hyphenation{Bool-ean corres-pond}

\title{Classical and Quantum Algorithms for the Boolean Satisfiability Problem}

\author{Carlos Barr\'{o}n-Romero
\thanks{Universidad Aut\'{o}noma Metropolitana, Unidad
 Azcapotzalco, Av. San Pablo No. 180, Col. Reynosa Tamaulipas,
 C.P. 02200. MEXICO.
cbarron@correo.azc.uam.mx
 }
}
\date{October 2, 2015}

\maketitle

\begin{abstract}

This paper presents a complete algorithmic study of the decision
Boolean Satisfiability Problem under the classical computation and
quantum computation theories. The paper depicts deterministic and
probabilistic algorithms, propositions of their properties and the
main result is that the problem has not an efficient algorithm (NP
is not P). Novel quantum algorithms and propositions depict that
the complexity by quantum computation approach for solving the
Boolean Satisfiability Problem or any NP problem is lineal time.

\end{abstract}



\pagestyle{myheadings} \thispagestyle{plain} \markboth{Carlos
Barr\'{o}n-Romero}{Classical and Quantum Algorithms for SAT}

\section{Introduction}


The complexity of algorithms have the classification NP-Soft
$\preccurlyeq $ NP-Hard. The problems in NP (Soft or Hard) are in
a classes, with two characteristics they have a verification
algorithm with polynomial complexity, and any problem in NP can be
translated between them.

There is no a book about algorithms, complexity or theory of
computation that it has not the description of the Boolean
Satisfiability Problem, named SAT. It is a classical problem and
one of the first to be in NP-Soft.

A Boolean variable only takes the values: $0$ (false) or $1$
(true). The logical operators are \textbf{not}: $\overline{x};$
\textbf{and}: $\wedge ,$ and \textbf{or}: $\vee .$

A SAT problem is a system Boolean formulas in  conjunctive normal
form over Boolean variables. Hereafter \SAT \ is a problem with
$n$ Boolean variables, and $m$ row Boolean formulas with at least
one Boolean variable. By example, SAT$(4,3)$:

\begin{equation*}
\begin{array}{ccccc}
\ \    & (x_{3}& \vee \ \overline{x}_{2}& & \vee \  x_{0}) \\
\wedge & (x_{3} & \vee \ x_{2}& \vee \ x_{1}) & \\
\wedge & ( \         &  \ \overline{x}_{2} & \vee \  x_{1} & \vee
\
x_{0}) \\
\wedge & (\  & & & \ \ \ x_{0}).%
\end{array}%
\end{equation*}

The assignation $x_{0}=1,x_{1}=0,x_{2}=1,$ and $x_{3}=1$ is a
solution because, such values satisfied the previous SAT$(4,3)$
but the assignation $x_{0}=0,x_{1}=0,x_{2}=0,x_{3}=0$ does not
satisfied it.

The problem is to determine, does \SAT\ have a solution? The
answer yes or no is the decision problem. Being skeptical, also a
specific solution is needed, the complexity of the verification is
polynomial.

Any \SAT \ can be translated into a set of ternary numbers. For
this $\Sigma =\left\{ 0,1,2\right\},$ and each row of \SAT \ maps
to a ternary number, with the convention: $\overline{x}_{i}$ to
$0$ (false), $x_{i}$ to $1$ ( true), and $2$ when the variable
$x_{i}$ is no present.

For example SAT$(5,3)$:
\begin{equation*}
\begin{array}{cccccc}
\ \ & ( \ x_{4} & \vee \  \overline{x}_{3} & & & \vee \  x_{0}) \\
\wedge & (\ \ \ \  & \ \ x_{3} & \vee \ x_{2} & & \ \ \ \ \  \ ) \\
\wedge & (\ \overline{x}_{4}& \vee \ x_{3} & \vee \  \overline{x}_{2} & \vee \ x_{1}
& \ \ \ \ \  \ ).%
\end{array}%
\end{equation*}

It is traduced to:
\begin{equation*}
\begin{array}{l}
10221 \\
21122 \\
01012.%
\end{array}%
\end{equation*}%

The assignment $x_{0}=1$, $x_{3}=1$ satisfies the previous
SAT$(5,3)$. It is not unique. Under this formulation, the search
space of SAT is $[0,3^{n}-1]$ where $n$ is the number of Boolean
variables.

Moreover, reciprocally the number $21221$ can be represented as a
Boolean formula. The previous SAT$(5,3)$ is modified as
SAT$(5,4)$:

\begin{equation*}
\begin{array}{cccccc}
\ \ & ( \ x_{4} & \vee \  \overline{x}_{3} & & & \vee \  x_{0}) \\
\wedge & (\  \ \ & \ \  x_{3} & \vee \ x_{2} & & \ \ \ \ \  \ ) \\
\wedge & (\overline{x}_{4}& \vee \ x_{3} & \vee \
\overline{x}_{2} & \vee \  x_{1} & \ \ \ \ \  \ ) \\
\wedge & (  \ \ \ &  \ \ x_{3} & & & \vee \  x_{0}).
\end{array}%
\end{equation*}

where $21221$ is translated into the last formula. Here, the
system is like a fixed point system, where $21221$ is a solution
into the set of numbers of SAT$(5,4)$. This point out that to look
for a solution is convenient to star with the numbers of \SAT, and
then continues with the numbers in $[0,3^{n}-1].$

The size of research space of SAT$(n,m)$ Boolean variables under
the ternary translation is $3^n$, this means: 1) the number of
different Boolean equations ($m$) could be from $1$ to $3^n$, 2)
It is drawback to analyze the row formulas in order to determine
properties or to rearrange their equations when $m$ is large. For
$m \approx 3^{n}$, $m$ is an exponential factor to take in
consideration to build knowledge for \SAT.

On the other hand, \SAT \ can be studied as a fixed point system,
and there is a simple subproblem  with equal Boolean formula´s
size \SSAT where the translated formulas is onto $\{0,1\}$. It has
the nice property to replace the huge search space from [0,
$3^{n}-1$] to $[0,$ $2^{n}-1]$.

Hereafter, the simple SAT is denoted by \SSAT  \ where $n$ is the
number of Boolean variables and $m$ is the number of rows the
problem with $n$ Boolean variables in each row .

An equivalent visual formulation is with $\blacksquare $ as 0, and
$\square $ as 1. Each row of a \SSAT\  uses the $n$ Boolean
variables to create a board, i.e. there is not missing variables.
The following boards have not a set of values in $\Sigma $ to
satisfy them:

\begin{equation*}
\begin{tabular}{|c|}
\hline
$x_{1}$ \\ \hline
$\blacksquare $ \\ \hline
$\square $ \\ \hline
\end{tabular}%
\ \ \
\begin{tabular}{|l|l|}
\hline
$x_{2}$ & $x_{1}$ \\ \hline
$\blacksquare $ & $\blacksquare $ \\ \hline
$\square $ & $\square $ \\ \hline
$\square $ & $\blacksquare $ \\ \hline
$\blacksquare $ & $\square $ \\ \hline
\end{tabular}%
\end{equation*}

I called unsatisfactory boards to the previous ones. It is clear
that they have not a solution because each number has its binary
complement, i.e., for each row image there is its complement
image. By example, the third row ($\square \blacksquare $) and
fourth row ($ \blacksquare \square $) are complementary.

To find an unsatisfactory board is like order the number and its
complement, by example: $000,$ $101,$ $110,$ $001,$ $010,$ $111,$
$011,$ and $100$ correspond to the unsatisfactory board:

\begin{equation*}
\begin{array}{c}
000 \\
111 \\
001 \\
110 \\
010 \\
101 \\
011 \\
100.%
\end{array}%
\end{equation*}

An algorithm to solve \SSAT \ by building an unsatisfactory board
is:

\begin{algorithm}
~\label{alg:SATBoard_1} \textbf{Input:} \SSAT.

\textbf{Output:} The answer if \SSAT \ has solution or not. $T$ is
an unsatisfactory board when \SSAT\ has not solution.

\textbf{Variables in memory}: $T[0:2^{n}-1]$=$-1$: array of binary
integer; $address$: integer; $ct=0$: integer; $k$: binary integer.

\begin{enumerate}
\item \textbf{while not end(}\SSAT \textbf{)}

\item \hspace{0.5cm} $k=b_{n-1}b_{n-2}\ldots b_{0}$=
\textbf{Translate to binary formula} (\SSAT);

\item \hspace{0.5cm} \textbf{if }$k.[b_{n-1}]$ \textbf{equal} $0$
\textbf{then}

\item \qquad \qquad $address=2\ast k.[b_{n-2}\ldots b_{0}];$

\item \qquad \textbf{else}

\item \qquad \qquad $address=2\ast (2^{n-1}-k.[b_{n-2}\ldots
b_{0}])-1;$

\item \hspace{0.5cm} \hspace{0.5cm} \textbf{end if}

\item \hspace{0.5cm} \textbf{if} T[$address$] \textbf{equal} $-1$
\textbf{then}

\item \hspace{0.5cm}\hspace{0.5cm} $ct=ct+1$;

\item \hspace{0.5cm}\hspace{0.5cm} $T[address]=k;$

\item \hspace{0.5cm} \textbf{end if}

\item \hspace{0.5cm}\textbf{if} $ct$ \textbf{equal} $2^{n}$
\textbf{then}

\item \hspace{0.5cm} \hspace{0.5cm} \textbf{output:} There is not
solution for \SSAT.

\item \hspace{0.5cm} \hspace{0.5cm} \textbf{stop}

\item \hspace{0.5cm} \textbf{end if} \item \textbf{end while}
\item \textbf{output:} \SSAT \ has a solution.
\end{enumerate}
\end{algorithm}

When \SSAT \ has not solution, table T is the witness because it
is an unsatisfactory board. On the other hand, when \SSAT \ has
solution the algorithm testify that exists a solution. The
complexity of the previous algorithm is $\mathbf{O}\left(
m\right)$. No matters if the rows of \SSAT \ are many or
duplicates or disordered, however it could finished around $2^n$
iterations.

It will take many unnecessary steps to use a sorting procedure but
Address Calculation Sorting (R. Singleton, 1956)~\cite{Agt:pena}.
To use a sorting procedure or to look for duplicate rows are not
convenient when $m\approx 2^{n}$. Singleton´s sorting uses direct
addressing into the search space $[0,2^n-1]$, and keeps the
complexity to the numbers of rows of \SSAT.

The resulting board contains in the odd rows $x=0b_{n}\ldots
b_{0}$ and in the next row $\overline{x}$. A solution of \SSAT\
is the first address of the rows of $T$ where there is not a pair
$x$ and $\overline{x}.$

In order to look for the solution of \SSAT, no previous knowledge
is assumed. It is no necessary to set in order the rows of \SSAT,
this algorithm determines for each row its address by addressing
directly to its board position. The visual inspection of $T$ point
out a solution of \SSAT, however a particular solution could be
difficult to find by inspection when $m$ is very large.

The algorithm~\ref{alg:SAT_2} \ in the next section provides the
answer and the solutions with the similar complexity.

\begin{proposition}
~\label{prop:NoSolSAT_binary} Let SSAT$(n,2^n)$ be where formulas
as the squares correspond to the $0$ to $2^{n}-1$ binary numbers.
Then it is an unsatisfactory board.
\begin{proof}
The binary strings of the values from $0$ to $2^l-1$ are all
possible assignation of values for the board. These strings
correspond to all combinations of $\Sigma^n$. It means that for
any possible assignation, there is the opposite Boolean formula
with value 0 and therefore SSAT$(n,2^n)$ has not solution.
\end{proof}
\end{proposition}

\begin{proposition}
~\label{prop:NoSolSAT} Given \SAT. There is not solution, if $L$
exists, where $L$ is any subset of rows, such that they are
isomorphic to an unsatisfactory board.
\begin{proof}
The subset $L$ satisfies the
proposition~\ref{prop:NoSolSAT_binary}. Therefore, it is not
possible to find satisfactory set of $n$ values for \SAT.
\end{proof}
\end{proposition}

Here, the last proposition depicts a necessary condition in order
to determine the existence of the solution for SAT. It is easy to
understand but it is highly complicated to determine the existence
of an unsatisfactory board in the general SAT. However, this point
out to focus studying the simple SAT with Boolean formulas where
each row has the same number of variables, i.e., \SSAT.

Hereafter, a more simply version of SAT is studied to map into
binary strings with the alphabet $\Sigma =\left\{ 0,1\right\} .$
Here, each row of \SSAT\  uses all the $n$ Boolean variables. This
allows to define the binary number problem of the translated row
formulas.
The binary number problem consists to find a binary number from $0$ to $%
2^{n}-1$ that has not its binary complement into the translated
values of the formulas of the given \SSAT. When the formulas of
the given \SSAT\  are different and $m=2^{n}$ by the proposition
\ref{prop:NoSolSAT_binary} there is not satisfactory assignment.

\begin{proposition}
~\label{prop:SolSAT_binary} Let \SSAT \ be with different row
formulas, and $m<2^{n}$.

There is a satisfactory assignation that correspond to a binary
string in $\Sigma^n$ as a number from $0$ to $2^{n}-1$.

\begin{proof}

Let  $s$ be any binary string that corresponds to a binary number
from $0$ to $2^l-1$, where $s$  has not a complement into the
translated formulas. Then $s$ coincide with at least one binary
digit of each binary number of the translated row formula.
Therefore, each row of the \SSAT\  is true.

\end{proof}
\end{proposition}

~\label{rem:WhereToLook} The previous proposition explains when
$s\in[0, 2^n-1]$ exists for \SSAT. Also,
proposition~\ref{prop:NoSolSAT_binary} states that if $m=2^n$ and
\SSAT has different rows, then there is not a solution. These are
necessary conditions for any \SSAT.

The complexity to determine such $s$ is depicted in the next
section.


\section{\SSAT\  has not properties for an efficient deterministic algorithm}

~\label{sc:SAT}

This section is devoted to \SSAT\  with each row formula has $n$
Boolean variables in descent order from $n-1$ to $0$. This allows
to translate each row to a unique binary string in $\Sigma^n$.

Recall, for SAT, I apply my technique: 1) to study general
problem, 2) to determine a simple reduction, and 3) to analyze
that there is no property for building an efficient algorithm for
the simple problem.

On the other hand, the simple reduction provides a simple version of SAT,
where for convenience all formulas have the same number of variables in a
given order. In any case, this simple version of SAT is for studying the
sections of complex SAT, and it is sufficient to prove that there is not
polynomial time algorithm for it.

The situation to solve \SSAT\ is subtle. Its number of rows could
be exponential, but no more than $2^n$. It is possible to consider
duplicate rows but this is not so important as to determine the
set $\mathcal{S}\subset \Sigma^n$, where $\mathcal{S}$ is the set
of the satisfactory assignations. On the other hand, $\Sigma^n$ is
the search space for any \SSAT. It corresponds to a regular
expression and it is easy to build it by a finite deterministic
automata (Kleene's Theorem).

\begin{proposition}
Let  $\Sigma={0,1}$ be an alphabet. Given a \text{SAT}$_{n\times
m}$, the set  $\mathcal{S} \subset \Sigma^n$ of the satisfactory
assignations is a regular expression.
\begin{proof}
$|\mathcal{S}|$ is a finite.
\end{proof}
\end{proposition}

It is trivial but from the computational point of view, the
construction and translation of strings associate to \SSAT  are
easy to build.

Solving \SSAT  could be easy if we have the binary numbers that
has not a complement in its translated rows. Also, because,
$|\Sigma^n|=2^n$ has exponential size, it is convenient to focus
in the information of \SSAT.

\SSAT\ can be transformed in a fixed point formulation. This
formulation is easy to understand.

\begin{figure}[tbp]
\centerline{\psfig{figure=\IMAGESPATH/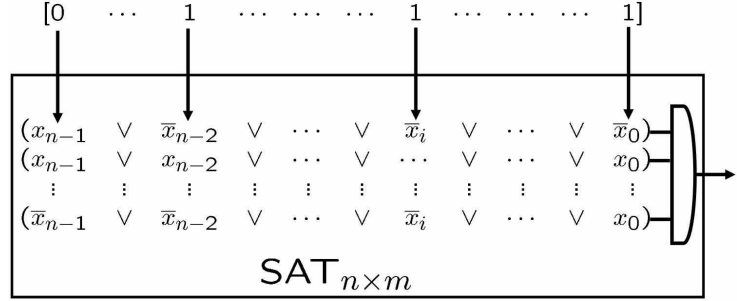, height=40mm}}~
\caption{\SSAT\ is a white of box containing a circuit of logical
gates where each row has the same number of Boolean variables.}
\label{fig:BoxSATnxm}
\end{figure}

\begin{enumerate}
\item An alphabet $\Sigma =\{0,1\}$.

\item Each  Boolean variable $x_{i}$ is mapped to $1$, and $\overline{x}%
_{j}$ to $0$.

\item A formula: $\overline{x}_{n-1}\vee \cdots x_{l}\vee \cdots
\vee \overline{x}_{1}\vee x_{0}$ can be transformed in its
corresponding binary string in $\Sigma ^{n}$; hereafter a binary
string is considered by its
value a binary number. By example $010\in \Sigma ^{3}$ is the binary number $%
10_{2}.$

\item If a $y\in $ $\Sigma ^{n}$, $\overline{y}$ is the complement
binary string. This is done bit a bit where $0$ is replaced by
$1,$ and $1$ by $0$.

\item \SSAT\ corresponds the set $M_{n\times m}$ of the $m$ binary
strings of its formulas.

$M_{n\times m}=\left\{%
\begin{array}{c}
s_{n-1}^{1}s_{n-2}^{1}\cdots s_{1}^{1}s_{0}^{1}, \\
s_{n-1}^{2}s_{n-2}^{2}\cdots s_{1}^{2}s_{0}^{2}, \\
\ldots , \\
s_{n-1}^{m}s_{n-2}^{m}\cdots s_{1}^{m}s_{0}^{m}%
\end{array}%
\right\}.$

Note that the number $s_{n-1}^{k}\cdots s_{1}^{k}s_{0}^{k}$
correspond to $k$ row of the \SSAT.

\item  \SSAT \ is a Boolean function. \SSAT$:\Sigma
^{n}\rightarrow \Sigma $. The argument is a binary string of $n$
bits. An important consideration is that the complexity of the
evaluation of this function is $\mathbf{O}(1)$. The
figure~\ref{fig:BoxSATnxm} depicts \SSAT\ as its logic circuit.

\item The decision problem SAT is equivalent to determine if \SSAT
\ is a blocked board or if exits $s\in[0,2^n]$ such that
\SSAT$(s)=1$.

\end{enumerate}

\begin{figure}[tbp]
\centerline{\psfig{figure=\IMAGESPATH/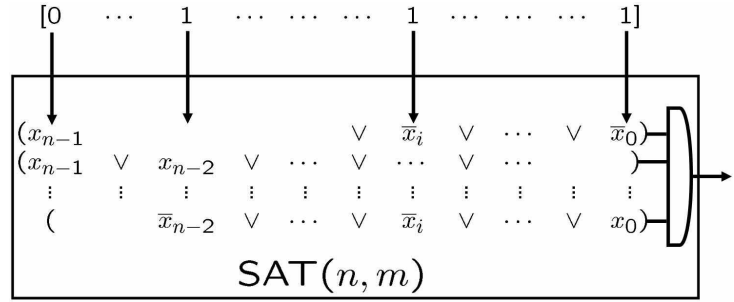,
height=40mm}}~ \caption{Final step of the algorithm
~\ref{alg:qSAT} at step 11 with the solution for \SAT.}
\label{fig:BoxqSATSol}
\end{figure}

~\label{rem:ComBoxSAT} The complexity of the evaluation of
\SSAT$(y=y_{n-1}y_{n-2}\cdots $ $y_{1}$ $y_{0})$ could be considered to be $%
\mathbf{O}(1)$. Instead of using a cycle, it is plausible to
consider that \SSAT is a circuit of logical gates. This is
depicted in figures~\ref{fig:BoxSATnxm} and \ref{fig:BoxqSATSol}.
Hereafter, \SSAT\ correspond to a logic circuit of "and", "or"
gates, and the complexity of the evaluation of the SAT is
$\mathbf{O}(1)$. The reason of this is to evaluate the SAT in an
appropriate time, one step, no matters if $m\approx 2^{n}$.

\begin{proposition}
~\label{prop:NumbSolSSAT}

Let \SSAT \ have different row formulas, and $m \leq 2^{n}$.

\begin{enumerate}
    \item The complexity to solve \SSAT\  is $\mathbf{O}(1).$
    \item Any subset of $\Sigma^n$ could be a solution for an appropriate
\SSAT.
\end{enumerate}

\begin{proof} \

\begin{enumerate}
    \item With the knowledge that the Boolean formulas are
    different, \SSAT\ has solution when $m <2^n$, i.e., it does not correspond to a blocked board.
    It has not
    solution when $m=2^n$, i.e., it is a blocked board.

    \item Any string of $\Sigma^n$ corresponds to a number in $[0,2^n-1]$.

 $\phi$ is the solution of a blocked board., i.e.,  for any
\SSAT \ with $m=2^n$.

For $m=2^n-1$, it is possible to build a \SSAT \  with only $x$ as
the solution. The blocked numbers $[0,2^n-1]$ $\setminus$ $\{x,
\overline{x}\}$ and $x$ are copied to M$_{n\times m}$. By
construction, \SSAT$(x)=1,$ i.e., it is unique, and belongs to
$\Sigma$.

 For $f$ different solutions. Let $x_1,\ldots,x_f$ be the given expected solutions.
 Build the set $C$ from the given solutions without any blocked
 pairs. Then the blocked numbers
$[0,2^n]$ $\setminus$ $\{y \in \Sigma^n |  x \in C, y=x \text{ or
} y= \overline{x}  \}$ and the numbers of $C$
 are put in M$_{n\times
m}$.
\end{enumerate}
\end{proof}
\end{proposition}

The proposition~\ref{prop:NumbSolSSAT}\ depicts the importance of
the knowledge. Here, "different rows formulas", it implies that
solving \SSAT \ is trivial. On the other hand, how much cost is to
determine or build such knowledge. People build algorithms based
in previous knowledge, many times without nothing at all.

The study of \SSAT \ is without any previous assumptions, it is a
white box with row Boolean formulas in conjunctive normal form.
Under this point of view, it is not possible to know if the rows
are different at first.

Let it be the set $\widehat{M}\left( \text{SAT}_{n\times m}\right)
=\{x\in \Sigma ^{n}|$ \SSAT$(x)=0\}$. It contains all the
complement binary string that they are no solution of \SSAT.

The complement of a set is defined as usual, and it is denoted by
$^{c}$.

$\widehat{M%
}\left( \text{SAT}_{n\times m}\right) ^{c}=$ $\{x\in \Sigma
^{n}|x\notin \widehat{M}\left( \text{SAT}_{n\times m}\right) \}.$
The context universal set is $\Sigma ^{n},$ and the empty set is
denoted by $\phi .$

\begin{proposition}
~\label{prop:SATSetSol} Any $x\in \widehat{M}\left(
\text{SAT}_{n\times m}\right) ^{c}$ is satisfactory assignment
of\break \SSAT.

\begin{proof}
It is immediately, $\widehat{M}\left( \text{SAT}_{n\times
m}\right) ^{c}=$ $\{x\in \Sigma ^{n}| \text{SAT}_{n\times m}(x) =
1 \}$.
\end{proof}
\end{proposition}

The numerical formulation provides immediately a necessary
condition for the existence of the solution of \SSAT: the set
$\widehat{M}\left( \text{SAT}_{n\times m}\right) ^{c}$ must be $\
$not empty. Here, we have three numerical binary sets
$M_{n\times m},$ $\widehat{M}\left( \text{SAT}_{n\times m}\right) $ and $%
\widehat{M}\left( \text{SAT}_{n\times m}\right) ^{c}.$

\begin{proposition}
Given \SSAT. If $M_{n\times m}\cap \widehat{M}\left( \text{SAT}%
_{n\times m}\right) ^{c}\neq \phi $. Then \SSAT\ is like fixed
point function$,i.e.,$ $\forall x\in M_{n\times m}\cap
\widehat{M}\left( \text{SAT}_{n\times m}\right) ^{c}$, $x\in
M_{n\times m}.$ On the other hand, If $M_{n\times m}\cap
\widehat{M}\left( \text{SAT}_{n\times m}\right) ^{c}=\phi ,$ then
for a $x\in \widehat{M}\left( \text{SAT}_{n\times m}\right) ^{c}$,
$M_{n\times m}\cup \left\{ x\right\} $ can be transformed into
\SSAT, adding the translation of $x$ as a row formula in \SSAT.
\end{proposition}

It is very expensive to analyze more than one time all the
formulas of \SSAT. When $m\approx 2^{n},$ any strategy for looking
properties has $m$ as a factor.

\begin{proposition}
~\label{prop:EvalMatchFixedPoint} If $y\in \Sigma ^{n}$, $%
y=y_{n-1}y_{n-2}\cdots y_{1}y_{0},$ then following strategies of
resolution of \SSAT\ are equivalent.

\begin{enumerate}

\item The evaluation of \SSAT$(y)$ as logic circuit.

\item~\label{stp:match} A matching procedure that consists verifying that each $%
y_{i}$ match at least one digit $s_{i}^{k}\in M_{n\times m},$ $\forall
k=1,\ldots ,m$.

\end{enumerate}

\begin{proof}
\SSAT$(y)=1$, it means that at least one variable of each row is
1, i.e., each $y_i,$ $i=1,\ldots,n$ for at least one bit, this
matches to 1 in $s^k_j$, $k=1,\ldots, m$.
\end{proof}
\end{proposition}

The evaluation strategies are equivalent but the computational
cost is not. The strategy~\ref{stp:match} implies at least $m
\cdot n$ iterations. This is a case for using each step of a cycle
to analyze each variable in a row formulas or to count how many
times a Boolean variable is used.

\begin{proposition}
An equivalent formulation of \SSAT\ is to look for a binary number
$x^{\ast }$ from $0$ to $2^{n}-1.$

\begin{enumerate}
\item If $x^{\ast }\in M_{n\times m}$ and $\overline{x}^{\ast
}\notin M_{n\times m}$ then \SSAT$(x^{\ast })=1.$

\item If $x^{\ast }\in M_{n\times m}$ and $\overline{x}^{\ast }\in
M_{n\times m} $ then \SSAT$(x^{\ast })=0.$ If $m<2^{n}-1$ then
$\exists y^{\ast} \in [0, 2^{n}-1]$ with $\overline{y}^{\ast
}\notin M_{n\times m} $ and \SSAT$(y^{\ast })=1.$

\item if 2), then $\exists$ SSAT$(n,m+1)$ such that 1) is fulfill.
\end{enumerate}

\begin{proof}  \

\begin{enumerate}
\item When $x^{\ast }\in M_{n\times m}$ and $\overline{x}^{\ast
}\notin M_{n\times m}$, this means that the corresponding formula
of $x^{\ast }$ is not blocked and for each Boolean formula of
\SSAT$(x^{\ast })$ at least one Boolean variable coincides with
one variable of $x^{\ast }.$  Therefore \SSAT$(x^{\ast })=1.$

\item I have,  $m<2^{n}-1$, then $\exists y^{\ast } \in[0,
2^{n}-1]$ with $\overline{y}^{\ast }\notin M_{n\times m}.$
Therefore, \SSAT$(y^\ast)=1$.

\item Adding the corresponding formula of $y^{\ast}$,
SSAT$(n,m+1)$ is obtained. By 1, the case is proved.
\end{enumerate}
\end{proof}
\end{proposition}

This approach is quite forward for verifying and getting a
solution for any \SSAT. By example, SSAT$(6,4)$ corresponds to the
set $M_{6\times 4}$:

\begin{equation*}
\begin{tabular}{|l|l|l|l|l|l|l|}
\hline
& $x_{5}=0$ & $x_{4}=0$ & $x_{3}=0$ & $x_{2}=0$ & $x_{1}=0$ & $x_{0}=0$ \\
\hline
& $\overline{x}_{5}\vee $ & $\overline{x}_{4}\vee $ & $\overline{x}_{3}\vee $
& $\overline{x}_{2}\vee $ & $\overline{x}_{1}\vee $ & $\overline{x}_{0})$ \\
\hline
$\wedge ($ & $\overline{x}_{5}\vee $ & $\overline{x}_{4}\vee $ & $\overline{x%
}_{3}\vee $ & $\overline{x}_{2}\vee $ & $\overline{x}_{1}\vee $ & $x_{0})$
\\ \hline
$\wedge ($ & $x_{5}\vee $ & $x_{4}\vee $ & $x_{3}\vee $ & $x_{2}\vee $ & $%
x_{1}\vee $ & $\overline{x}_{0})$ \\ \hline
$\wedge ($ & $\overline{x}_{5}\vee $ & $x_{4}\vee $ & $x_{3}\vee $ & $%
\overline{x}_{2}\vee $ & $x_{1}\vee $ & $x_{0})$ \\ \hline
\end{tabular}%
\text{ \ }%
\end{equation*}

\begin{equation*}
\begin{tabular}{|l|l|l|l|l|l|}
\hline
$x_{5}$ & $x_{4}$ & $x_{3}$ & $x_{2}$ & $x_{1}$ & $x_{0}$ \\ \hline
$0$ & $0$ & $0$ & $0$ & $0$ & $0$ \\ \hline
$0$ & $0$ & $0$ & $0$ & $0$ & $1$ \\ \hline
$1$ & $1$ & $1$ & $1$ & $1$ & $0$ \\ \hline
$0$ & $1$ & $1$ & $0$ & $1$ & $1$ \\ \hline
\end{tabular}%
\text{.}
\end{equation*}

The first table depicts that SSAT$(6,4)(y=000000)=1$. The second
table depicts the set $M_{6\times 4}$ as an array of binary
numbers. The assignation $y$ corresponds to first row of
$M_{6\times 4}.$ At least one digit of $y$ coincides with each
number of
M$_{n\times m}$, the Boolean formulas of SAT$(6,4).$ Finally, $y$ $=$ $%
000000$ can be interpreted as the satisfied assignment $x_{5}=0,$
$x_{4}=0,$ $x_{3}=0,$ $x_{2}=0,$ $x_{1}=0,$ and $x_{0}=0.$

\begin{proposition}
~\label{prop:SATFixedPoint2} Given a \SSAT, there is a binary
number $y\in M_{n\times m}$ such that $\overline{y}\notin
M_{n\times m}$ then $y$ is fixed point for \SSAT\ or
SSAT$(n,m+1)(y)$, where SSAT$(n,m+1)$ is \SSAT\ with adding the
corresponding formula of $y.$
\begin{proof}
This result follows from the previous proposition.
\end{proof}
\end{proposition}

\SSAT can be used as an array of $m$ indexed Boolean formulas. In
fact, the previous proposition gives an interpretation of the
\SSAT \ as a type fixed point problem. For convenience, without
before exploring the formulas the SAT, my strategy is to look each
formula, and to keep information in a Boolean array of the
formulas of SAT by its binary number as an index for the array. At
this point, the resolution \SSAT \  is equivalent to look for a
binary number $x$ such that \SSAT$\left( x\right) =1$. The
strategy is to use the binary number representation of the
formulas of \SSAT \  in M$_{n\times m}.$

A computable algorithm for solving \SSAT\  is:

\begin{algorithm}~\label{alg:SAT_2} \textbf{Input:} \SSAT.

\textbf{Output:} $k:$ integer, such that \SSAT$(k)=1$ or SSAT$%
(n,m+1)(k)=1$ or \SSAT\  has not solution.

\textbf{Variables in memory}: $T[0: 2^{n}-1]$: array of double linked structure $previous$, $next$: integer; $ct$:=$0$ :
integer; $first=0$: integer; $last=2^{n}-1$: integer;

\begin{enumerate}
\item \textbf{while not end(}\SSAT\textbf{)}

\item \hspace{0.5cm} $k$ $=$ \textbf{Translate to binary formula}
(\SSAT);

\item \hspace{0.5cm} \textbf{if } $T[k]$.next \textbf{not equal} $-1$ \textbf{then}

\item \hspace{0.5cm} \hspace{0.5cm} \textbf{if} \SSAT$(k)$
\textbf{equal} 1 \textbf{then}

\item \hspace{0.5cm} \hspace{0.5cm} \hspace{0.5cm}
\textbf{output:} $k$ is a solution for \SSAT;

\item \hspace{0.5cm} \hspace{0.5cm} \hspace{0.5cm} \textbf{stop};

\item \qquad \qquad \textbf{else} \%\ Update the links of $T$

\item~\label{stepLinkI} \hspace{0.5cm} \qquad \qquad
$T[T[k]$.previous].next=$T[k]$.next;

\item \qquad \qquad \qquad $T[T[k]$.next].previous =
$T[k]$.previous;

\item \qquad \qquad \qquad \textbf{if} $k$ \textbf{equal} $first$ \textbf{then}

\item \qquad \qquad \qquad \qquad $first$ := $T[k]$.next;
\item \qquad \qquad \qquad \textbf{end if}

\item \qquad \qquad \qquad \textbf{if} $k$ \textbf{equal} $last$ \textbf{then}

\item \qquad \qquad \qquad \qquad $last$ := $T[k]$.previous;
\item \qquad \qquad \qquad \textbf{end if}

\item \qquad \qquad \qquad $T[k]$.next=$-1$;

\item \qquad \qquad \qquad $T[k]$.previous =$-1;$

\item \hspace{0.5cm} \qquad \qquad
$T[T[\overline{k}]$.previous].next=$T[ \overline{k}]$.next;

\item \qquad \qquad \qquad $T[T[\overline{k}]$.next].previous =
$T[\overline{k}]$.previous;

\item \qquad \qquad \qquad \textbf{if} $\overline{k}$ \textbf{equal} $first$ \textbf{then}

\item \qquad \qquad \qquad \qquad $first$ := $T[\overline{k}]$.next;
\item \qquad \qquad \qquad \textbf{end if}

\item \qquad \qquad \qquad \textbf{if} $\overline{k}$ \textbf{equal} $last$ \textbf{then}

\item \qquad \qquad \qquad \qquad $last$ := $T[\overline{k}]$.previous;
\item \qquad \qquad \qquad \textbf{end if}

\item \qquad \qquad \qquad $T[\overline{k}]$.next=$-1$;

\item~\label{stepLinkF} \qquad \qquad \qquad
$T[\overline{k}]$.previous =$-1;$

\item \hspace{0.5cm} $\qquad \qquad ct$ := $ct$ +$2$;

\item \hspace{0.5cm} \hspace{0.5cm} \textbf{end if}

\item \hspace{0.5cm} \textbf{end if}

\item \hspace{0.5cm} \textbf{if} ct \textbf{equal} $2^{n}$
\textbf{then}

\item \hspace{0.5cm} \hspace{0.5cm} \textbf{output:} There is not
solution for \SSAT;

\item \hspace{0.5cm} \hspace{0.5cm} \textbf{stop};

\item \hspace{0.5cm} \textbf{end if}

\item \textbf{end while}

\item $k = first$;

\item~\label{stp:chngSATnmm} \textbf{output:} $k$ a the solution
for SSAT$(n,m+1)$ adding the corresponding Boolean formula $k$;

\item \textbf{stop};
\end{enumerate}
\end{algorithm}

It is not necessary to use $m$, the algorithm works while
there is a row to analyze.

Line~\ref{stp:chngSATnmm} changes the input \SSAT\ to
SSAT$(n,m+1)$. This is for making SSAT$(n,m+1)$ as point fixed
problem, i.e., if the algorithm runs again, the solutions is
already a formula in SSAT$(n,m+1).$

$T$ is an special vector array of links. It has a memory cost of
$2^{n+1}$. Updating the links has a low fixed cost as it is
depicted in lines~\ref{stepLinkI} to~\ref{stepLinkF}. The array
$T$ allows to design the previous algorithm in efficient way, to
computes all the solutions, but the draw back is the memory cost.

Assuming that \SSAT\  is a logic circuit (see remark~\ref%
{rem:ComBoxSAT}) the complexity of its evaluation is
$\mathbf{O}(1)$. This is a non trivial assumption that encloses
that the Boolean formulas of a given \SSAT\  are given but not
analyzed previously. In fact, the formulas of the given problem
could be repeated or disordered.

If a solution is not in M$_{n\times m}$, the complexity of the
previous algorithm is a cycle of size $m$ plus 1, i.e.,
$\mathbf{O}(m)$. But the worst case scenario is when $m\approx
2^{n}$, i.e., $\mathbf{O}(m)$ $=$ $\mathbf{O}(2^{n})$. If the
Boolean formulas are in separated of their complement, the
complexity is $\mathbf{O}(m/2)$. Nevertheless, with duplicated
formulas and without solution the algorithm is computable, i.e.,
the number of iteration is always bounded, with no duplicated
formulas, it is at most $2^n$.

For the modified SSAT$(n,m+1)$, inverting the cycle "for", i.e.,
\textbf{for} i:= m+1 \textbf{downto} 1, the complexity is
$\mathbf{O}(1)$. The knowledge pays off but it is not a property
to exploit to reduce the complexity for any \SSAT. It is a
posteriori property to build SSAT$(n,m+1)$ and to take advantage
of it as a fixed point type of problem.

The next algorithm computes, in similar way, the set
$\mathcal{S}\subset \Sigma^n$ of the solutions, i.e., the
knowledge of \SSAT. When $|\mathcal{S}|$ is small or zero, to
determine all the solutions or no solution has the same complexity
$\mathbf{O}(2^n).$

\begin{algorithm}~\label{alg:SATKnow_3}\textbf{Input:} \SSAT.

\textbf{Output:} $T:$ List of binary numbers such that, $x\in T$,
\SSAT(x)=1, or \SSAT \ has not solution and $T$ is empty.

\textbf{Variables in memory}: $T[0:2^{n-1}]$: list as an array of
integers, double link structure $previous$, $next$ : integer;
$ct$:=$0$ : integer;  $first=0$: integer; $last=2^{n}-1$: integer;

\begin{enumerate}
\item \textbf{while not end(}\SSAT\textbf{)}

\item \hspace{0.5cm} $k$ $=$ \textbf{Translate to binary formula}
(\SSAT);

 \item \hspace{0.5cm} \textbf{if} $T[k]$.previous
\textbf{not equal} $-1$ \textbf{ or } $T[k]$.next \textbf{not
equal} $-1$ \textbf{then}

\item \hspace{0.5cm} \hspace{0.5cm} \textbf{if} \SSAT$(k)$
\textbf{not equal} 1 \textbf{then} \%\ Update the links of $T$

\item \hspace{0.5cm} \qquad \qquad $T[T[k]$.previous$]$.next $=T[k]$.next;

\item \qquad \qquad \qquad $T[T[k]$.next$]$.previous $=T[k]$.previous;

\item \qquad \qquad \qquad \textbf{if} $k$ \textbf{equal} $first$ \textbf{then}

\item \qquad \qquad \qquad \qquad $first$ := $T[k]$.next;
\item \qquad \qquad \qquad \textbf{end if}

\item \qquad \qquad \qquad \textbf{if} $k$ \textbf{equal} $last$ \textbf{then}

\item \qquad \qquad \qquad \qquad $last$ := $T[k]$.previous;
\item \qquad \qquad \qquad \textbf{end if}

\item \qquad \qquad \qquad $T[k].next=-1$;

\item \qquad \qquad \qquad $T[k]$.previous =$-1;$

\item \hspace{0.5cm} \qquad \qquad $T[T[\overline{k}]$.previous$]$.next $=T[%
\overline{k}]$.next;

\item \qquad \qquad \qquad $T[T[\overline{k}]$.next$]$.previous $=T[%
\overline{k}]$.previous;

\item \qquad \qquad \qquad \textbf{if} $\overline{k}$ \textbf{equal} $first$ \textbf{then}

\item \qquad \qquad \qquad \qquad $first$ := $T[\overline{k}]$.next;
\item \qquad \qquad \qquad \textbf{end if}

\item \qquad \qquad \qquad \textbf{if} $\overline{k}$ \textbf{equal} $last$ \textbf{then}

\item \qquad \qquad \qquad \qquad $last$ := $T[\overline{k}]$.previous;
\item \qquad \qquad \qquad \textbf{end if}

\item \qquad \qquad \qquad $T[\overline{k}]$.next $=-1$;

\item \qquad \qquad \qquad $T[\overline{k}]$.previous $=-1;$

\item \hspace{0.5cm} $\qquad \qquad ct:=ct+2;$

\item \hspace{0.5cm} \hspace{0.5cm} \textbf{end if}

\item \hspace{0.5cm} \textbf{end if}

\item \hspace{0.5cm} \textbf{if} $ct$ \textbf{equal} $2^{n}$
\textbf{then}

\item \hspace{0.5cm} \hspace{0.5cm} \textbf{output:} There is not solution for SAT$%
_{n\times m}$;

\item \hspace{0.5cm}  \hspace{0.5cm} \textbf{stop};

\item \hspace{0.5cm} \textbf{end if}

\item \textbf{end while}

\item \textbf{stop}
\end{enumerate}
\end{algorithm}

It is not necessary to use $m$, the algorithm works while
there is a row to analyze.

For the extreme cases, when $n$ is large and $m\approx2^n$ with
no duplicates formulas, the cost of solving \SSAT \ or building
$T$ is similar, $\mathbf{O}(m).$ There is a high probability that
without any knowledge of the positions of the formulas, the
algorithm~\ref{alg:SAT_2} executes $m$ steps.

 At the end of the previous algorithm, the knowledge of
\SSAT is the set $\mathcal{S}$. Before of exploring \SSAT, there
are not plausible binary numbers associated with its solution in
one step.

Now, let $\mathbb{K}_{n\times m}$ be knowledge of the solutions.

\begin{proposition}~\label{prop:SATKnow} Given \SSAT.

\begin{enumerate}
\item $\mathbb{K}_{n\times m}=\widehat{M}\left(
\text{SAT}_{n\times m}\right) ^{c}=\mathcal{S}.$

\item The cost of building $\mathbb{K}_{n\times m}$ is
$\mathbf{O}(m).$

\item It is $\mathbf{O}=1$ by solving SSAT$(n,m+1)$ from
$\mathbb{K}_{n\times m}.$

\end{enumerate}

\begin{proof} \

\begin{enumerate}

    \item The algorithm~\ref{alg:SATKnow_3} builds $\widehat{M}\left( \text{SAT}_{n\times
m}\right) ^{c}$ = $\mathbb{K}_{n\times m}.$ Also, by definition,
$\mathcal{S}$ $=$ $\widehat{M}\left( \text{SAT}_{n\times m}\right)
^{c}.$

    \item The algorithm~\ref{alg:SATKnow_3} has only a cycle of size $m$.

\item Any $s\in \widehat{M}\left( \text{SAT}_{n\times m}\right)
^{c}$ solves SSAT$(n,m+1)$ in one iteration.
\end{enumerate}
\end{proof}
\end{proposition}

The proposition depicts the condition for solving efficiently
\SSAT, which is $\mathbb{K}_{n\times m}$ the knowledge associated
with the specific given \SSAT. After analyzing the formulas of
\SSAT, we have $\widehat{M}\left( \text{SAT}(n,m)\right)^{c}$ $=$
$\mathbb{K}_{n\times m}$ but no before.

The two previous algorithm use the array $T$. It combines index
array and double links, the drawback is the amount of memory
needed to update the links, but its cost is $\mathbf{O}(1)$.

The next tables depict that inserting and erasing binary numbers
into $T$ (this structure correspond to $\mathcal{S}$) only
consists on updating links with fixed complexity $\mathbf{O}(8)$
(number of the link assignations).

\begin{center}
\begin{tabular}{|c|c|c|}
\hline
\multicolumn{3}{|c|} {$first=0$; $last=7;$} \\ \hline
\multicolumn{3}{|c|}{$T$} \\ \hline $i$ & $previous$ &
$next$ \\ \hline $0$ & $-1$ & $1$ \\ \hline $1$ & $0$ & $2$ \\
\hline $2$ & $1$ & $3$ \\ \hline $3$ & $2$ & $4$ \\ \hline $4$ & $3$ & $5$ \\
\hline $5$ & $4$ & $6$
\\ \hline $6$ & $5$ & $7$ \\ \hline $7$ & $6$ & $-1$ \\ \hline
\end{tabular}
\end{center}

By example, let SSAT$(3,2)$ be

$$
\begin{array}{cccc}
\ \ &(\ x_{2} &\vee \ \overline{x}_{1} & \vee \ x_{0}) \\
\wedge & (\ x_{2} & \vee \  \overline{x}_{1} & \vee \ \overline{x}_{0}).%
\end{array}%
$$

The resulting array $T$ with $\mathbb{K}_{3\times 2}$ after
executing algorithm~\ref{alg:SATKnow_3} is

\begin{center}

\begin{tabular}{|c|c|c|}
\hline \multicolumn{3}{|c|} {$first=0$; $last=7;$} \\ \hline
\multicolumn{3}{|c|}{$T$} \\ \hline $i$ & $previous$ &
$next$ \\ \hline $0$ & $-1$ & $1$ \\ \hline $1$ & $0$ & $4$ \\
\hline $2$ & $-1$ & $-1$ \\ \hline $3$ & $-1$ & $-1$ \\ \hline
$4$ & $1$ & $5$ \\
\hline $5$ & $4$ & $6$
\\ \hline $6$ & $5$ & $7$ \\ \hline $7$ & $6$ & $-1$ \\ \hline
\end{tabular}%
\end{center}

The knowledge for a given \SSAT\ depends of exploring it. Before,
to explore \SSAT\ there are not properties nor binary numbers
associated with $\mathbb{K}_{n\times m}$. By properties, I mean,
the algorithm~\ref{alg:SATKnow_3} computes $\mathcal{S}$, now it
is possible to state the properties of the all number in
$\mathcal{S}$, by example, they could bye twin prime numbers.

Having $\mathbb{K}_{n\times m}$, the knowledge of \SSAT \ for any
numbers $x$ of the list $T$, the complexity for verifying
\SSAT$(x)=1$ is  $\mathbf{O}(1)$.

It is trivial to solve SAT when knowledge is given or can be created in an
efficient way. This results of this section are related to my article ~\cite%
{arXiv:Barron2010}, where I stated that NP problems need to look
for its solution in an search space using a Turing Machine for
General Assign Problem of size $n$ (GAP$n$) : "It is a TM the
appropriate computational model for a simple algorithm to explore
at full the GAP$n$'s research space or a reduced research space of
it". As trivial as
it sound, pickup a solution for \SSAT\ depends of exploring its $%
m $ Boolean formulas.

The question if there exists an efficient algorithm for any \SSAT\
now can be answered. The algorithm~\ref{alg:SATKnow_3} is
technologically
implausible because the amount of memory needed. However, building $\mathbb{K%
}_{n\times m}$ provides a very efficient telephone algorithm (see
proposition 8, and remarks 5 in~\cite{arXiv:Barron2005}) where
succeed is guaranteed for any $s$ $\in $ $\mathbb{K}_{n\times m}$.
On the other hand, following ~\cite{arXiv:Barron2005}) there are
tree possibilities exhaustive algorithm (exploring all the
searching space), scout algorithm (previous knowledge or heuristic
facilities for searching in the search space), and wizard
algorithm (using necessary and sufficient properties of the
problem).

The study of NP problems depicts that only for an special type of
problem exists properties allowing to build an ad-hoc efficient
algorithms, but these properties can not be generalized for any
GAP or any member of class NP.

The Boolean formulas of \SSAT\ correspond to binary numbers in
disorder (assuming an order of the Boolean variables as binary
digits). The algorithm~\ref{alg:SAT_2} has a complexity of the
size of $m$ (the numbers of formulas of SAT).It is not worth to
considered sorting algorithm because complexity increase by an
exponential factor $m\approx 2^n)$ when $m$ is very large.

Without exploring and doing nothing, for any \SSAT, the set
M$_{n\times m}$ can be an interpretation without the cost of
translation. The associated M$_{n\times m}$ can be consider an
arbitrary set of numbers. The numbers in M$_{n\times m}$ have not
relation or property between them to point out what are the
satisfied binary number in $\Sigma^n$. The numbers of M$_{n\times
m}$, as binary strings have the property to belong to the
translation of \SSAT\ but $\mathcal{S} \cap$ M$_{n\times m}$
$=\phi$.

In fact one or many numbers could be solution or not one, but also
it is not easy to pick up the solution in the set $[0,2^{n}-1]$
without knowledge.

Hereafter, the Boolean formulas of \SSAT \  are considered a
translation to binary numbers in disorder and without any
correlation between them.

\begin{proposition}
With algorithm~\ref{alg:SAT_2}:

\begin{enumerate}
\item A solution for \SSAT\ is efficient when $m << $ $2^{n}.$

\item A solution for \SSAT\ is not efficient when $m \approx $
$2^{n}.$
\end{enumerate}

\begin{proof}
It follows from algorithm~\ref{alg:SAT_2}.
\end{proof}
\end{proposition}

\begin{proposition}
The complexity to determine if \SSAT has solution, or to build the
knowledge or to find the set $\mathcal{S}$ of \SSAT is the same
and it is around $\mathbf{O}(m)$.

\begin{proof}
It follows from algorithms~\ref{alg:SATBoard_1},~\ref{alg:SAT_2},
and~\ref{alg:SATKnow_3}.
\end{proof}
\end{proposition}

The previous proposition states the complexity of the
deterministic way to solve \SSAT without no prior knowledge. It is
important to note that there are not iterations or previous steps
to study \SSAT. Any of the
algorithms~\ref{alg:SATBoard_1},~\ref{alg:SAT_2},
and~\ref{alg:SATKnow_3} face the problem without any assumptions of
what formulas, or order, or structure could have it. For they, it
is a like a circuit in a white box as a file of logic formulas.

The next section studies \SSAT \ from the probabilistic point of
view.


\section{Probabilistic algorithm for \SSAT}

In this section the research space $[0,2^{n}-1]$ is considered as
a collection of objects with the same probability to be selected.
Hereafter, \Prob \ stands for a probability function.

\begin{proposition}
Let \SSAT\ have $n$ large with no duplicates rows.

\begin{enumerate}

\item $x\in [0,2^{n}-1]$ has the same priori probability to be
selected, i.e, uniform probability.

\item \Prob$(\mathcal{S})$ $=|\mathcal{S}|/2^n$ is fixed.

\end{enumerate}

\begin{proof} \

By the proposition~\ref{prop:NumbSolSSAT} any arbitrary set
$\mathcal{S}\subset \Sigma^n$ ($|\Sigma^n|=2^n$) could be the set
of solutions of an appropriate \SSAT.

\begin{enumerate}
\item  Without previous knowledge or reviewing the formulas of
\SSAT\ any number could be selected with \Prob$(x)=1/2^n.$

\item Any $x\in \mathcal{S}$ is the solution and  any $y\in
\mathcal{S}^c$ is not. This is fixed for a given \SSAT. Therefore
$1=$\Prob$(\mathcal{S})$ $+$ \Prob$(\mathcal{S}^c).$ Moreover,
\Prob$(\mathcal{S})$ $=|\mathcal{S}|/2^n.$
\end{enumerate}

\end{proof}
\end{proposition}

The uniform probability for selecting any $x\in [0,2^{n}-1]$ is
\Prob$(\{x\})=1/2^n.$

\SSAT is solved using the properties:
\begin{enumerate}
    \item \SSAT is a function. Its formulas are disordered.
    \item The problem to solve is to determine if exist or not a binary number $x
    \in[0,2^n-1],$ such that \SSAT$(x)=1.$
    \item $\forall x\in[0,2^n-1]$, such that \SSAT$(x)=0$ then
    \SSAT$(\overline{x})=0.$
    \item $n$ and $m$ are large.
    \item Without any previous knowledge,  any $x\in
[0,2^{n}-1]$ has the same priori probability to be selected, i.e,
uniform probability.
    \item  $m$ is arbitrary large, including the case $m>2^n$.
    This means, \SSAT \ could have duplicate
    formulas.

\item \Prob$(\mathcal{S})$ $=|\mathcal{S}|/2^n$ is fixed.

    \item After testing $f$ binary numbers $x_1,x_2,\ldots,x_f$, such that \SSAT$(x_i)=0$,
    the failure set is $F=\{ x_1, \ldots, x_f, \} $.
The probability of the selection of the rest candidates for
solving \SSAT  slightly increase, i.e.,  the posteriori
probability of the candidates for solving \SSAT\ of any $x\in
[0,2^{n}-1] \setminus F$ is equal to $1/\left( 2^{n}-\left\vert
F\right\vert \right) = 1/\left( 2^{n}- f \right) \approx 1/2^{n}.$
\end{enumerate}

\begin{proposition}~\label{prop:probsel}
Let $n$ be large.

\begin{enumerate}

\item When $m\approx 2^n$ or $m>2^n$, the probability for
selecting a solution after $f$ failures ($f<<2^n$) is
insignificant.

 \item  A solution for \SSAT\ is efficient when $m
<< $ $2^{n}.$

\item  A solution for \SSAT\ is not efficient when $m \approx $
$2^{n}.$
\end{enumerate}

\begin{proof} \

\begin{enumerate}
\item Let $P(\text{So})=|\mathcal{S}|/2^{2n}$ be the probability
of selecting a solution $s\in \mathcal{S}$, where $\mathcal{S}$ is
the set of the solutions. Assuming $m$ huge, \SSAT\ has many
different rows (also, it is possible to have duplicates rows),
therefore the set $|\mathcal{S}|$ is very small. Let
$P(\text{Se}(f))$ $=$ $1/\left( 2^{n}-f \right)$ be the
probability after $f$ failures. Then the probability
$P(\text{Se}(f) \cap \text{So})$ $=$ $1/\left( 2^{n}-f \right)
\cdot |\mathcal{S}|/2^n \approx |\mathcal{S}|/2^{2n}$ $\approx$
$0.$

 \item When $m \ll 2^n,$  \SSAT \ has a high probability that its rows are not
 blocked, i.e., $|\mathcal{S}|$ $\approx 2^n$. Therefore, the
probability for solving \SSAT, $P(\text{Se})$ $=$
$|\mathcal{S}|/2^n$ is almost $1$, i.e., many numbers in
$[0,2^n-1]$ are solutions. It is fast to pick $x$ $\in$
$[0,2^n-1]$ $\cap$ $\mathcal{S}$.

\item In this case, there are only few numbers to solve \SSAT \
many rows of \SSAT \ are blocked, i.e., $|\mathcal{S}|$ $\ll 2^n$.
$P(\text{Se}(f) \cap \text{So})$ $\approx |\mathcal{S}|/2^{2n}
\approx 0$. It is insignificant and take a long time to find a
solution.
\end{enumerate}

\end{proof}
\end{proposition}

In order to find a solution for SAT, there are a probabilistic
algorithms rather than the previous deterministic algorithms. By
example, a probabilistic algorithm for \SSAT\ is:

\begin{algorithm}~\label{alg_prob:SAT_4} \textbf{Input:} \SSAT.

\textbf{Output:} $k:$ integer, such that \SSAT$(k)=1$ or
SSAT$(n,m+1)(k)=1$ or \SSAT \  is not satisfied and all entries of
$T$ are 1.

\textbf{Variables in memory}: $T[0:2^{n-1}]=0$ of Boolean; $ct$:=0
: integer;

\begin{enumerate}
\item \textbf{while (1)}

\item~\label{stp:selection} \hspace{0.5cm} \textbf{Select uniform
randomly} $k \in [0,2^{n}-1] \setminus \{ i \, |\,  T[i] =1\}$;

\item \hspace{0.5cm} \textbf{if} \SSAT$(k)$ \textbf{equals} 1
\textbf{then}

\item \hspace{0.5cm} \hspace{0.5cm} \textbf{output:} $k$ is the
solution for \SSAT.

\item \hspace{0.5cm} \hspace{0.5cm} \textbf{stop}

\item \hspace{0.5cm} \textbf{end if}

\item \hspace{0.5cm} \textbf{if} $T[k]$ \textbf{not equal} $1$
\textbf{then}

\item \hspace{0.5cm} \hspace{0.5cm} $T[k]:=1$;

\item \hspace{0.5cm} \hspace{0.5cm} $ct := ct + 1$;

\item \hspace{0.5cm} \hspace{0.5cm} \textbf{if} $ct$
\textbf{equal} $2^{n}$ \textbf{then}

\item \hspace{0.5cm} \hspace{0.5cm} \hspace{0.5cm}
\textbf{output:} There is not solution for \SSAT.

\item \hspace{0.5cm} \hspace{0.5cm} \hspace{0.5cm} \textbf{stop}

\item \hspace{0.5cm} \hspace{0.5cm} \textbf{end if}

\item \textbf{end while}
\end{enumerate}
\end{algorithm}

This algorithm is computable. It always finishes with the answer
for any \SSAT. In each cycle, the probability of the selection of
the candidates is slightly increased and one binary number is
omitted from the list of candidates but the selection is still
probabilistic uniform for the remain candidates. It is
$1/(2^n-k)$ after $k$ cycles.

\begin{proposition} The algorithm~\ref{alg_prob:SAT_4}:

\begin{enumerate}
\item is efficient for finding a solution for \SSAT\ when $m << $
$2^{n}.$

\item is not efficient for finding a solution or to determine if
\SSAT has a solution when $m \approx $ $2^{n}$ or $m \geq 2^n$.
\end{enumerate}

\begin{proof} \

\begin{enumerate}

\item When $m << $ $2^{n}$, it implies, \Prob(\SSAT$(k)=1$)
$\approx 1$ for many $k$. The algorithm is highly probable to
solve \SSAT \ in short time.

\item When $m \approx $ $2^{n}$, it is highly probable that many
selected uniform randomly number in $[0,2^n-1]$ are blocked, i,e.,
\Prob(\SSAT $(k)=0)$ $\approx 1$ for many $k$ of the step 2. The
algorithm~\ref{alg_prob:SAT_4} needs to test a huge amount of
numbers for increasing the probability for selecting a solution.
An extreme case is with only few binary numbers as the solution,
but the worst case is when \SSAT \ is a blocked board, in this
case the algorithm executes $2^{n}$ steps. Even with duplicates
formulas, i.e., when $m \geq 2^n$, the algorithm takes at less
$2^{n}$ steps to determine
 no solution.
\end{enumerate}
\end{proof}
\end{proposition}

Finally,

\begin{proposition}
~\label{prop:SATnxm} \SSAT\ has not property or heuristic to build
an efficient algorithm.

\begin{proof}

For any \SSAT, the translation of its formulas to binary numbers
is implicit without cost by assuming the following implicit rules:
$x_i$ is 1, and $\overline{x}_i$ is $0,$ and the order of the
variables in each row is the same. M$_{n\times m}$ is the set of
binary numbers, it could be in disorder, and without any previous
knowledge, also, there are not correlation between the binary
strings numbers in it. The only property between number is
\SSAT$(x)=0$ if and only if \SSAT$(\overline{x})=0.$ But, this
property does not increase the probability for be the solution
after $f\ll 2^n$ failures by prop~\ref{prop:probsel}.

If a property or a heuristic exists such that an arbitrary
probabilistic algorithm quickly determines when \SSAT \ has or not
a solution then even in the extreme case (when there is not
solution, i.e., when the formulas of \SSAT \ correspond to a
blocked board), the number of steps could at  less
$2^{n-1}$ for the algorithms~\ref{alg:SAT_2} and~\ref{alg:SATKnow_3} or
$2^{n}$ for the algorithms~\ref{alg:SATBoard_1} and~\ref{alg_prob:SAT_4}.

For the safe of the argumentation let us considerer $n$ and $m$
large, $m \approx 2^n$ or $m \gg 2^n$. Also, the solution of \SSAT
could be any binary number or none.

With $m \gg 2^n$, the deterministic algorithms like the algorithms
~\ref{alg:SATBoard_1},~\ref{alg:SAT_2}, and~\ref{alg:SATKnow_3} are not
efficient, $\mathbf{O}(2^n)$ .

If an algorithm similar to the  algorithm~\ref{alg_prob:SAT_4} finds
very quickly $x^\ast\in[0,2^n-1]$ such that \SSAT$(x^\ast)=1$ then
 the selection of the candidates
(step~\ref{stp:selection}) has not uniform probability.

Using such property, any arbitrary subset of numbers must have it,
otherwise, the algorithm is not efficient for any \SSAT. But, this
property alters the prior probability of the uniform distribution.
Such characteristic or property implies that any natural number is
related to each other with no uniform probability at priory for be
selected.

This means, that such property is in the intersection of all
properties for all natural numbers. Also it is no related to the
value, because, the intersection of natural classes under the
modulo prime number is empty.

Moreover, it point outs efficiently to the solution of \SSAT, it
means that this property  is inherently to any number to make it
not equally probable for its selection. Also, the binary numbers
of M$_{n \times m}$ correspond to an arbitrary arrangement of the
Boolean variables of \SSAT, so in a different arrangement of the
positions for the Boolean variables, any number has inherently not
uniform probability for selection. On the other hand, for the
extreme case when the formulas of \SSAT \ correspond to a blocked
board, the property implies that no solution exits with the
numbers of steps
 $\ll 2^{n-1}$, i.e., the answer is found without reviewing all
 the formulas of \SSAT!
\end{proof}
\end{proposition}

It is an absurd that such property exits, because it changes the
probability of the uniform distribution of similar objects without
previous knowledge. Without an inspection of the formulas of \SSAT
when $n \gg 1$, and $m=2^n-1$, the unique solution has $1/2^n$ as
the priory probability of be the solution. After $k$ cycles, with
the information of the failed candidates, the probability slightly
grows to $1/(2^n-2k),$ i.e., the probability of the uniform
selection taking in consideration the failed candidates does not
grow exponentially but lineally. Therefore, any probabilistic
algorithm must take a lot of time to determine the solution of
\SSAT.

\begin{proposition}
~\label{prop:NPnP} NP has not property or heuristic to build an efficient
algorithm.

\begin{proof}
Let X be a problem in NP. PH${_\text{X}}$ is the set of properties
or heuristics for building an efficient algorithm for problem X.
$$\bigcap_\text{X} \text{PH}{_\text{X}}=\phi$$
by the previous proposition.
\end{proof}
\end{proposition}

\begin{proposition}
~\label{prop:Lower_boundSAT} A lower bound for the complexity of
\SSAT for the extreme case when there is no solution or only one
solution is $2^{n-1}-1$. Therefore, $\mathbf{O}(2^{n-1})$
$\preccurlyeq $ NP-Soft $\preccurlyeq $ NP-Hard.

\begin{proof}
For the case when there is only one solution, the probabilistic
algorithm~\ref{alg_prob:SAT_4} after $k$ failed numbers has a
probability equals $1/(2^n-k)$. The number of tries to get a
probability equals to $1/2$ is $k=2^{n}-2$. On the other hand,
with the algorithm~\ref{alg:SAT_2} for the case when there is no
solution the number of iterations could be at less $2^{n-1}$.

No matter if the algorithm is deterministic or probabilistic a
lower bound for the complexity of \SSAT \ is
$\mathbf{O}(2^{n-1})$. This implies, $\mathbf{O}(2^{n-1})$
$\preccurlyeq $ NP-Soft $\preccurlyeq $ NP-Hard.
\end{proof}
\end{proposition}

\begin{figure}[tbp]
\centerline{\psfig{figure=\IMAGESPATH/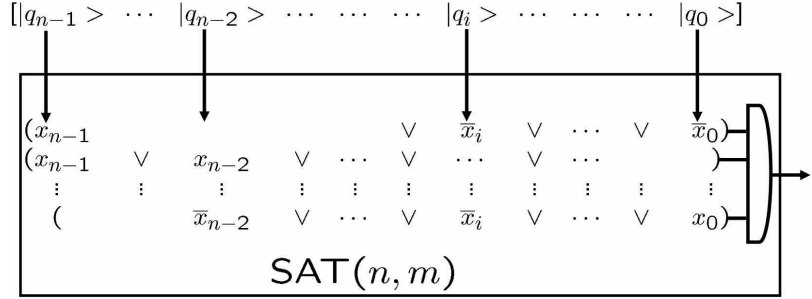, height=40mm}}~
\caption{Quantum Boolean variables input \SAT \ for the
algorithm~\ref{alg:qSAT} at step 1.} \label{fig:BoxqSAT}
\end{figure}

\begin{figure}[tbp]
\centerline{\psfig{figure=\IMAGESPATH/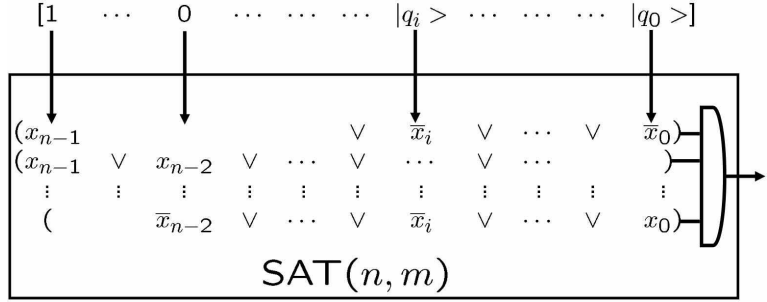,
height=40mm}}~ \caption{Boolean values, and quantum Boolean
variables input \SAT \ for the algorithm~\ref{alg:qSAT} at step
7.} \label{fig:Boxq_i_SAT}
\end{figure}

\section{Quantum computation}

Grover's Algorithm can be adapted to look for the binary number
that it solves \SSAT, adapting the function $C$ of the clauses of
SAT as Grover depicts in~\cite{arXiv:grover1996}. Using this
adaptation, the Grover's Algorithm is $\mathbf{O}(2^{n/2})$. This
is not good but it is better than $\mathbf{O}(2^n).$
In~\cite{siamc:Shor1997}, there is survey of the quantum potential
and complexity and the Shor's algorithm for prime factorization
and other discrete algorithms.

Here, the quantum computation approach is used to formulate a
novel algorithm based on quantum hardware for the general SAT.
Similar to the idea of Jozsa [1992] and Berthiaume and Brassard
[1992, 1994] of a random number generator with a Feynman~\cite{
addison:Feynman1996} reachable, and reversible quantum circuit
approach.

The main idea is to adapt the algorithm~\ref{alg:SAT_2} for using
quantum variables for uniformly exploring $[0,2^{n}-1]$ states at
the same time. The following proposition states how the quantum
Boolean variables behave:

\begin{proposition}
$n$ quantum Boolean Variables, uniformly explore $\{0,1\}^n$.
\begin{proof}
If they do not. Then there is a priori property that it can
explain why. But this means that they have not the random behavior
of the quantum phenomenon related.
\end{proof}
\end{proposition}

In the algorithm~\ref{alg:SAT_2}, the cycle \textbf{for} i:=$0$ to
$2^{n}-1$ is replaced by the input of $n$ quantum binary
variables.  Instead of doing iterations, the coupling of the
quantum variables with circuit of \SSAT \ reaches its reversible
and stable state in one operation. Figure~\ref{fig:BoxqSAT}
depicts quantum variables instead of the cycle \textbf{for} i:=$0$
to $2^{n}-1$. This approach is based to the couple quantum
hardware with a reversible quantum circuit of SAT. The only two
outcomes 0 or 1 are found after the quantum circuit reach it
stable and reversible state.

Here, the problem to solve is the general SAT, i.e., \SAT \ where
$n$ is the number of Boolean variables and its rows Boolean
formulas could have from one to $n$ Boolean variables. The
property to build this quantum hardware-software depends to couple
the input quantum variables with the Boolean circuit of \SAT \ for
interacting to reach a reversible and stable solution after the
quantum exploration of the search space. The assumption is that
$n$ quantum variables have the $2^{n}$ Boolean values as input of
\SAT. After a while or maybe instantaneous, the couple reach the
stable and reversible state, then the output of \SAT \ only have
two outcomes. When the outcome is 0, there is not solution.
Otherwise, a solution is computed bit by bit.

It is important to note that the complexity of solving \SAT \ is
independent of $m,$ the number of Boolean formulas.

The following algorithm depicts how to solve efficiently \SAT \ by
quantum computation.

\begin{algorithm}~\label{alg:qSAT}

\textbf{Input:} \SAT.

\textbf{Output:} $x$ such that \SAT$(x)=1$ or \SAT \ has not
solution, where $x[0:n-1]:$ \textbf{array of Boolean variables; }

\textbf{Variables in memory}: $q[0 :n-1]$: \textbf{Array of
quantum Boolean variables}; $v$: \textbf{integer.}

\begin{enumerate}
\item \textbf{if} \SAT$(q_{n},q_{n-1},\ldots ,q_{0})$
\textbf{equal} 0 \textbf{then}

\item \qquad \textbf{output:} There is not solution for \SAT.

\item \qquad\ \textbf{stop}

\item \textbf{end if}

\item \textbf{for }$v=n-2$ \textbf{downto} $0$

\item \qquad $x_{v+1}=0;$

\item \hspace{0.5cm} \textbf{if} \SAT$(x_{n-1},\ldots
,x_{v+1},q_{v},\ldots ,q_{0})$ \textbf{equals} 0 \textbf{then}

\item \qquad \qquad $x_{v+1}=1;$

\item \qquad \textbf{end if}

\item \textbf{end for}

\item \textbf{output:} $x=[x_{n-1},x_{n-1},\ldots ,x_{0}]$ is the
solution for \SAT.

\item \textbf{stop}
\end{enumerate}
\end{algorithm}

The first step is the answer to question if SAT has or not
solution. The rest of the steps build a solution by uniformly
exploring $\{0,1\}^v,$ $v=n-2,\ldots,0$ quantum Boolean variables.
It is crucial to compute a solution, as a witness to verify by
direct evaluation that such $x$ satisfies \SAT$(x)=1.$

Figure ~\ref{fig:Boxq_i_SAT} depicts the substitution of the
quantum variables by the corresponding Boolean values that
satisfies \SAT \  at the step $i$. Here, there are not a tree of
alternatives, there are only two choices for each Boolean variable
when the the quantum variables are been substituting.

If the assumption of the coupling quantum variables with a \SAT \
works, then the complexity of the previous algorithm is
$\mathbf{O}(1)$ to answer the decision SAT. But $\mathbf{O}(n)$ is
for building a solution $x$. With $x$ the verification that
\SAT$(x)=1$ is easy and straight forward.

Moreover, because of the equivalence of the class NP, any \SAT \
or NP problem under quantum computation has the same complexity!
The article~\cite{arXiv:Barron2010} depicts that the search space
of GAP is finite and numerable, therefore a similar a coupling
between quantum Boolean variables and the cycles can be used for
any NP.

On the other hand, the next algorithm verifies no solution for
SAT.
\begin{algorithm}~\label{alg:qVerSAT} \textbf{Input:} \SAT.

\textbf{Output:} \SAT is or not consistent with no solution;

\textbf{Variables in memory}: $q[0 :n-1]$: \textbf{Array of
quantum Boolean variables}; $v$: \textbf{integer.}

\begin{enumerate}
\item \textbf{if} \SAT$(q_{n},q_{n-1},\ldots ,q_{0})$
\textbf{equal} 0 \textbf{then}

 \item \textbf{for }$v=n-1$ \textbf{downto} $0$

\item \hspace{0.5cm} \textbf{if} \SAT$(q_{n-1},  \ldots
,0,q_{v},\ldots ,q_{0})$ \textbf{equals} 0 \textbf{or}

\hspace{0.5cm} \textbf{\ \ } \SAT$(q_{n-1},  \ldots
,1,q_{v},\ldots ,q_{0})$ \textbf{equals}  1 \textbf{then}

\item~\label{step:inconsistent} \hspace{0.5cm} \hspace{0.5cm}
\textbf{output:} No solution is inconsistent for \SAT.

\item \hspace{0.5cm} \hspace{0.5cm} \textbf{stop}

\item \hspace{0.5cm} \textbf{end if}

\item \textbf{end for}

\item \textbf{output:} No solution is consistent for \SAT.

\item \textbf{stop}

\end{enumerate}
\end{algorithm}

The previous algorithm has complexity $\mathbf{O}(n).$ It states
an important proposition.

\begin{proposition}~\label{prop:quantumToQuantum}
A quantum algorithm for uniformly exploring $\{0,1\}^n$ can be
only verified by a similar quantum algorithm in short time.
\begin{proof}
It follows from the propositions of the previous section that
quantum computational approach can review the search space
$\{0,1\}^n$ but traditional algorithms take exponential number of
iterations, i.e.,$\mathbf{O}(2^{n-1}).$
\end{proof}
\end{proposition}

To my knowledge, under the quantum theory the outcome of the
step~\ref{step:inconsistent} is impossible but human error or
failure in the construction of the couple quantum hardware with
the logic circuit.

\section*{Conclusions and future work}
~\label{sc:conclusions and future work}

Heuristic techniques, using previous knowledge do not provide
reducibility (see 6 in ~\cite{arXiv:Barron2010}) for all the NP
problems. The lack of a common property for defining an efficient
algorithm took a long way. My research focused in the verification
of the solutions in polynomial time. I point out that there is not
an efficient verification algorithm for solving a NP Hard problem
(GAP is a minimization problem). I develop a reduction method,
here it is applied to the SAT.

The classical SAT (a NP decision problem), where formulas have any
number of Boolean variables, is reduced to the simple version
\SSAT. This allows to focus in the characteristics and properties
for solving the simple \SSAT \ with the reduction of the search
space from $\{0,1,2\}^n$ to $\{0,1\}^n$, and the number problem
formulation for \SSAT. The proposition~\ref{prop:SATnxm} states
that there is not a property for building an efficient algorithm
for \SSAT. It is immediately an efficient algorithm does not exist
for any SAT or NP.

For the future, I am interesting in reviewing the computational
models, languages and theory of computation under the quantum
computational approach. Quantum computation is opening new
perceptions, paradigms and technological applications. Today or
near future, it is possible no programming at all, for building a
very complicate computational system will be to design and grow a
huge complex crystalline structure of a huge massive quantum logic
circuit. It is possible that the only way to test this type of
design is by the quantum computational approach as it is stated by
proposition~\ref{prop:quantumToQuantum}.

Finally, \SSAT \ supports and proves that there is not a property
to reduce the complexity of the worst case for a decision NP
problem, i.e., NP $\neq $ P. On the other hand, a coupling of
quantum variables with \SAT \ provides the novel
algorithm~\ref{alg:qSAT}, which it states: any NP problem has
lineal complexity for its solution under an appropriate quantum
computational design.


\end{document}